\documentclass[12pt,a4paper]{article}
\usepackage{epsfig}
\begin{document}


\title{RADIODETECTION OF COSMIC RAY EXTENSIVE AIR SHOWERS : 
PRESENT STATUS OF THE CODALEMA EXPERIMENT.}
\maketitle
\begin{center}
\author{$^{a)}$D.ARDOUIN\footnote{daniel.ardouin@sciences.univ-nantes.fr}
,  $^{a)}$BELLETOILE A., $^{a)}$CHARRIER D., $^{a)}$DALLIER R., 
 $^{b)}$DENIS L., $^{d)}$ESCHSTRUTH P., 
 $^{a)}$GOUSSET T., $^{a)}$HADDAD F.,
 $^{a)}$LAMBLIN J., $^{a)}$LAUTRIDOU P., $^{c)}$LECACHEUX A., 
 $^{d)}$MONNIER-RAGAIGNE D., $^{a)}$RAHMANI A.,\ $^{a)}$RAVEL O.\\
$^{a)}$SUBATECH, La Chantrerie, 4 rue Alfred Kastler, BP 20722, 44307 Nantes-cedex 3\\
$^{b)}$Observatoire de Paris - Station de radioastronomie, 18330 Nan\c{c}ay \\
$^{c)}$LESIA, Observatoire de Paris, Section de Meudon, 5 place Jules Janssen, 92190 MEUDON\\
$^{d)}$LAL, Universit\'e Paris-Sud, B\^atiment 200, BP 34, F-91898 Orsay cedex}
\end{center} 


\begin{abstract}
The CODALEMA  experiment uses 6 large frequency bandwidth antennas
 of the Nan\c cay Radio Observatory Decametric Array (France).
In a first configuration, one antenna narrowed band filtered acting as trigger, 
 with a 4 $\sigma$ threshold above sky background-level, was used to 
tag any radio transient in coincidence on the antenna array.  
Recently, the addition of 4 particle detectors allowed us 
to observe cosmic ray events in coincidence with antennas.
\end{abstract}

{\bf keywords:}ultra high energy cosmic rays ; radiodetection.


\section{The CODALEMA experiment.}
We present the characteristics and performances 
of a demonstrative experiment devoted to the observation of ultra high- energy
 cosmic rays extensive air showers using  a radiodetection technique. The 
 CODALEMA (COsmic ray Detection Array with Logarithmic Electromagnetic Antennas)
 experiment was set up at the Nan\c cay Radio Observatory in 2003.
 It uses 6 of the 144 log-periodic
 antennas (in the 1-100~MHz frequency band for CODALEMA) constituting the
 DecAMetric array (DAM) \cite{Dam}.

\begin{figure}[h]
\hfill
\begin{minipage}[t]{.45\textwidth}
   \centerline{\epsfig{file=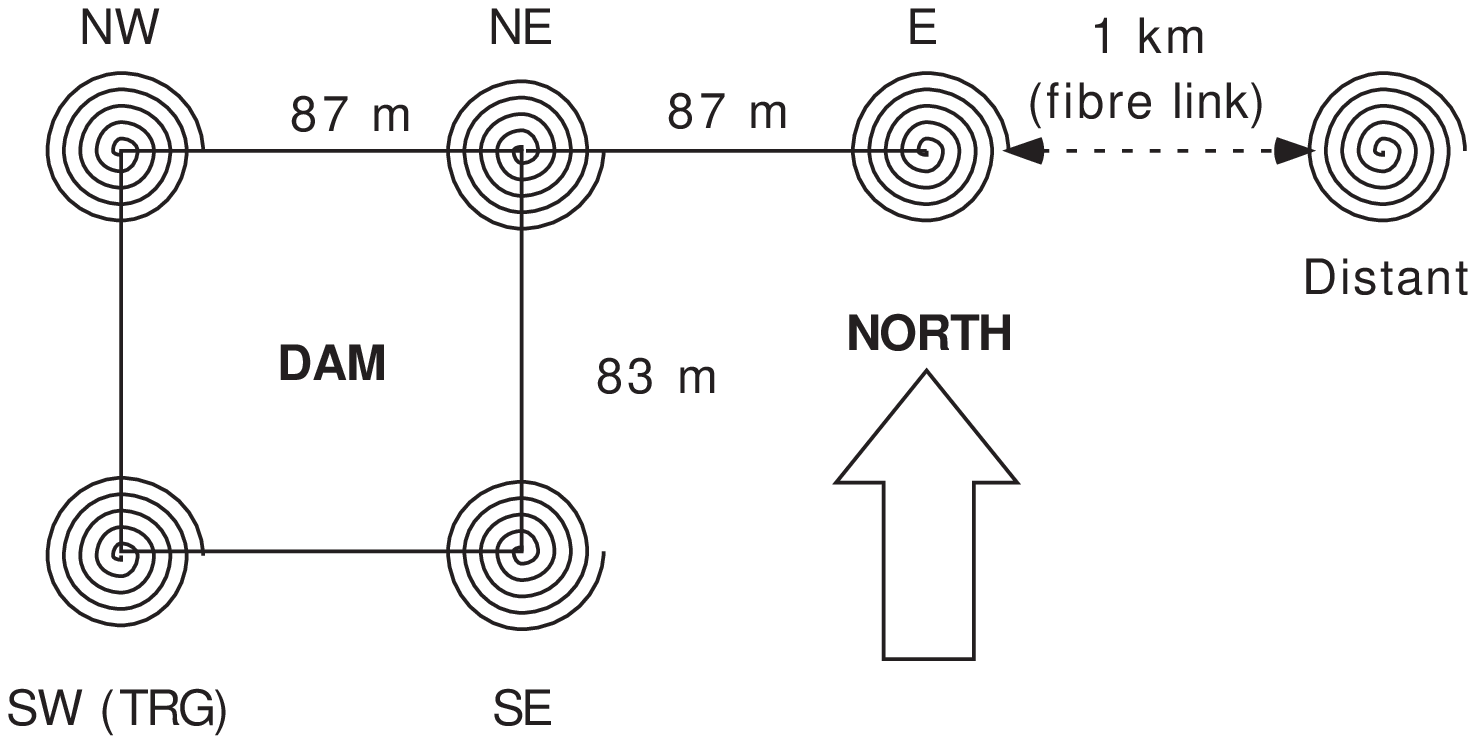,width=7cm}}
   \vspace*{8pt}
      \caption{First CODALEMA setup: the SW antenna acted as a trigger.\label{fig:setup}}
\end{minipage}
\hfill
\begin{minipage}[t]{.45\textwidth}
  \centerline{\epsfig{file=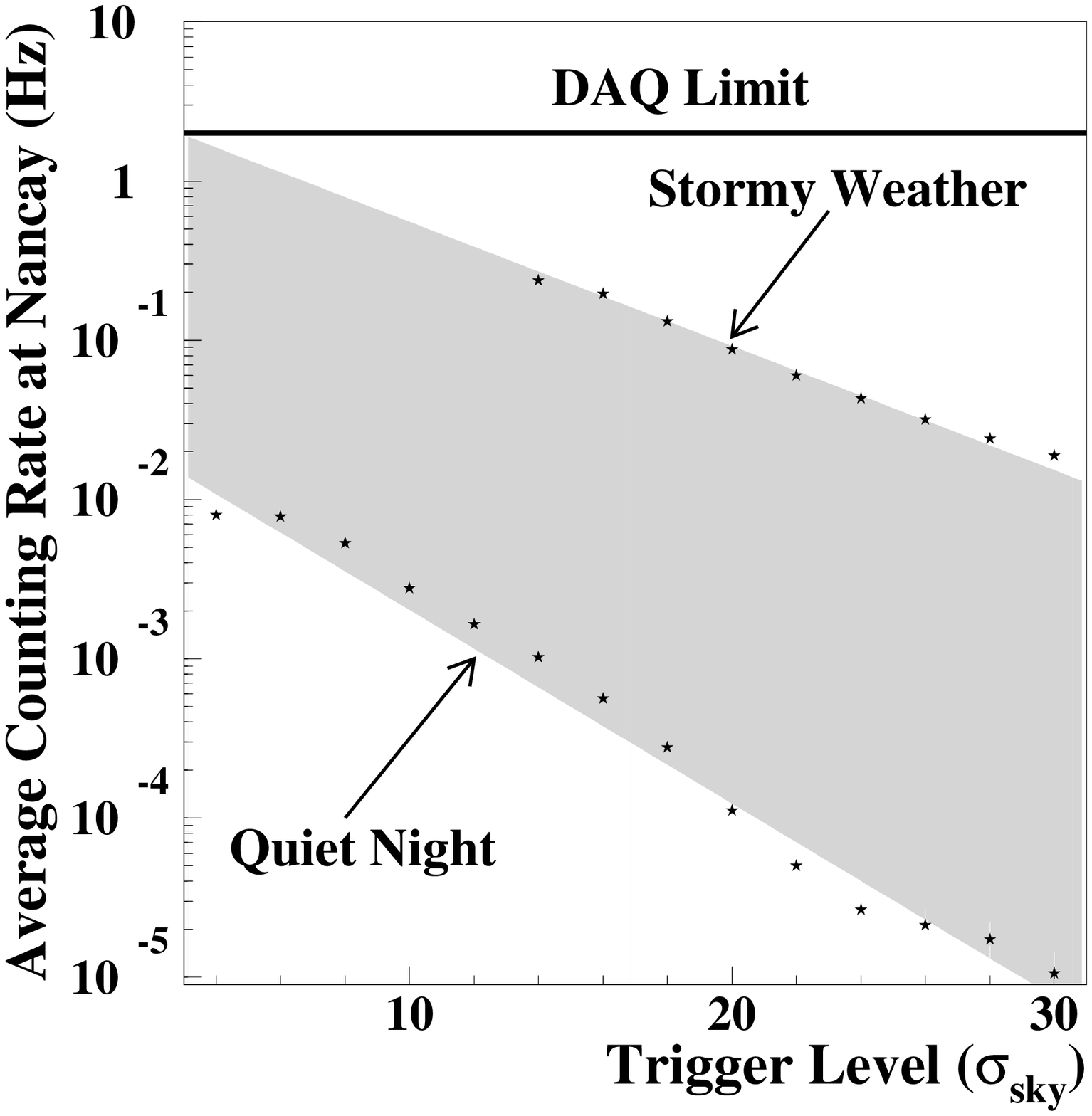,width=4.5cm}}
  \vspace*{8pt}
    \caption{The shaded area 
  corresponds to the measured counting rate. The lower limit has been measured during quiet
 night runs whereas the
   upper limit corresponds to stormy
  weather. \label{fig:triggerlevel}}
\end{minipage}
  \end{figure}

In the first period of observation \cite{Ravel}, the setup (see Fig. \ref{fig:setup}) was self-triggered using
 one devoted antenna: its signal was filtered in an appropriate noise-free
 frequency band (33-65~MHz) chosen after an exhaustive
 study in the observed local noise frequency spectrum,  before entering the ADC. 
The wide band waveform signals (1 - 100 MHz) of the other antennas were registered
 when a voltage threshold was reached on the trigger antenna. 
The trigger threshold was set at 4~$\sigma_{sky}$ ($\sigma_{sky}$: the rms sky background noise),
leading to  an electric field sensitivity of 4~${\mu}$V/m .

On figure \ref{fig:triggerlevel} the evolution of the average counting rate at Nancay is 
presented as a function of the trigger level expressed in unit of $\sigma_{sky}$.
 The counting rate evolves greatly with the anthropic
 activities in the vicinity of the station of Nancay and with the weather conditions. 

Except for the trigger antenna, transient signals on the antennas were hidden by 
radio transmitters signals. Consequently, a numerical passband filter (same as trigger frequency band)
 was applied, offline, in order to observe coincidences involving several antennas \cite{Dallier}. 
Using the position and the timing differences between antennas, it was also possible to perform   
 the trajectory reconstruction of the electromagnetic plane wave using a triangulation techniques
 across the array \cite{Arnaud}. This level of analysis enables us to bring in light several
 cosmic ray air shower candidates.

\section{Coincident particle data: Preliminary results}
\begin{figure}[h]
\hfill
\begin{minipage}[t]{.45\textwidth}
\centerline{\epsfig{file=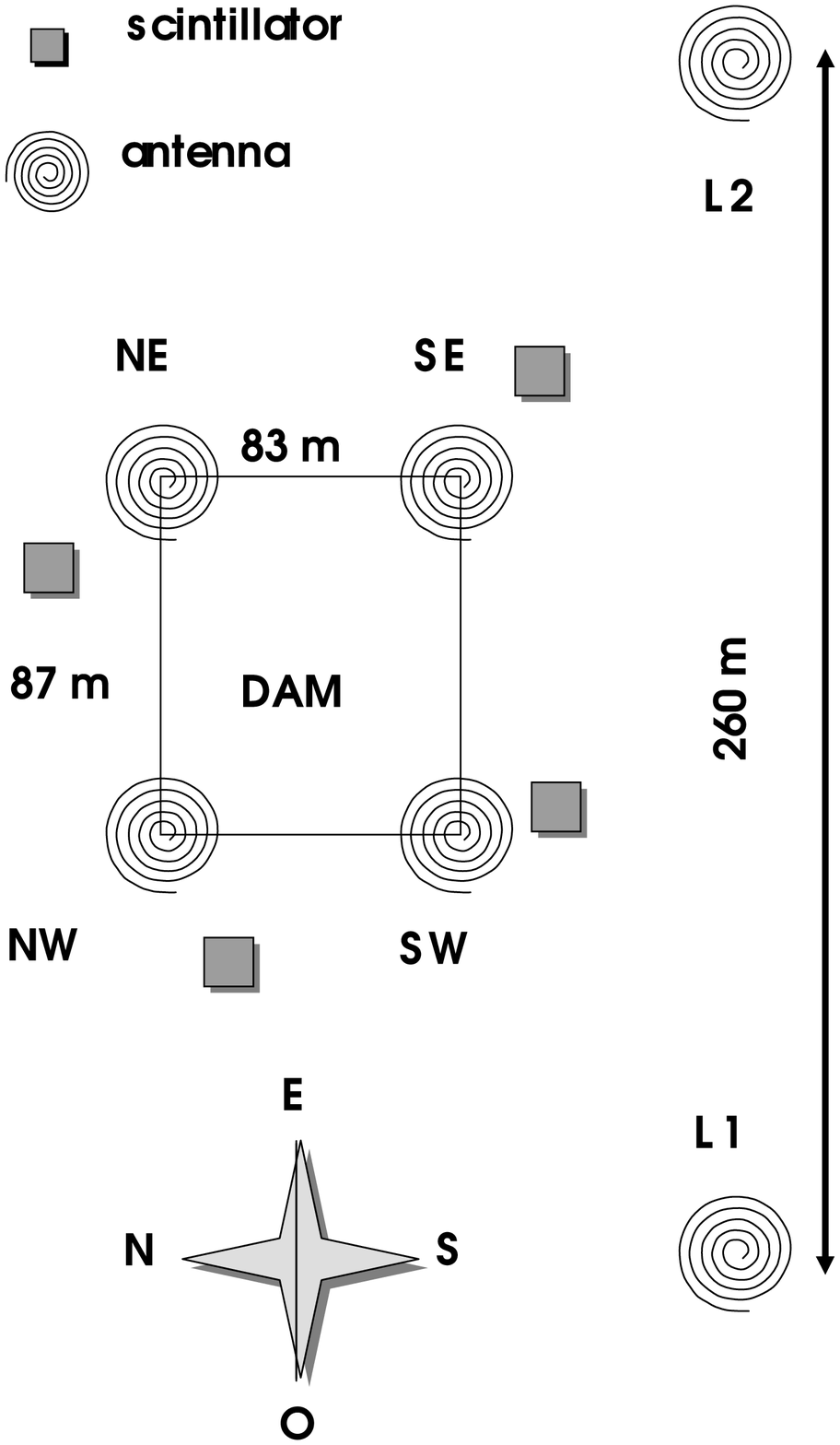,width=5cm}}
\vspace*{8pt}
\caption{Actual CODALEMA setup: the particle detectors act as a trigger.\label{fig:setup2}}
\end{minipage}
\hfill
\begin{minipage}[t]{.45\textwidth}
\centerline{\epsfig{file=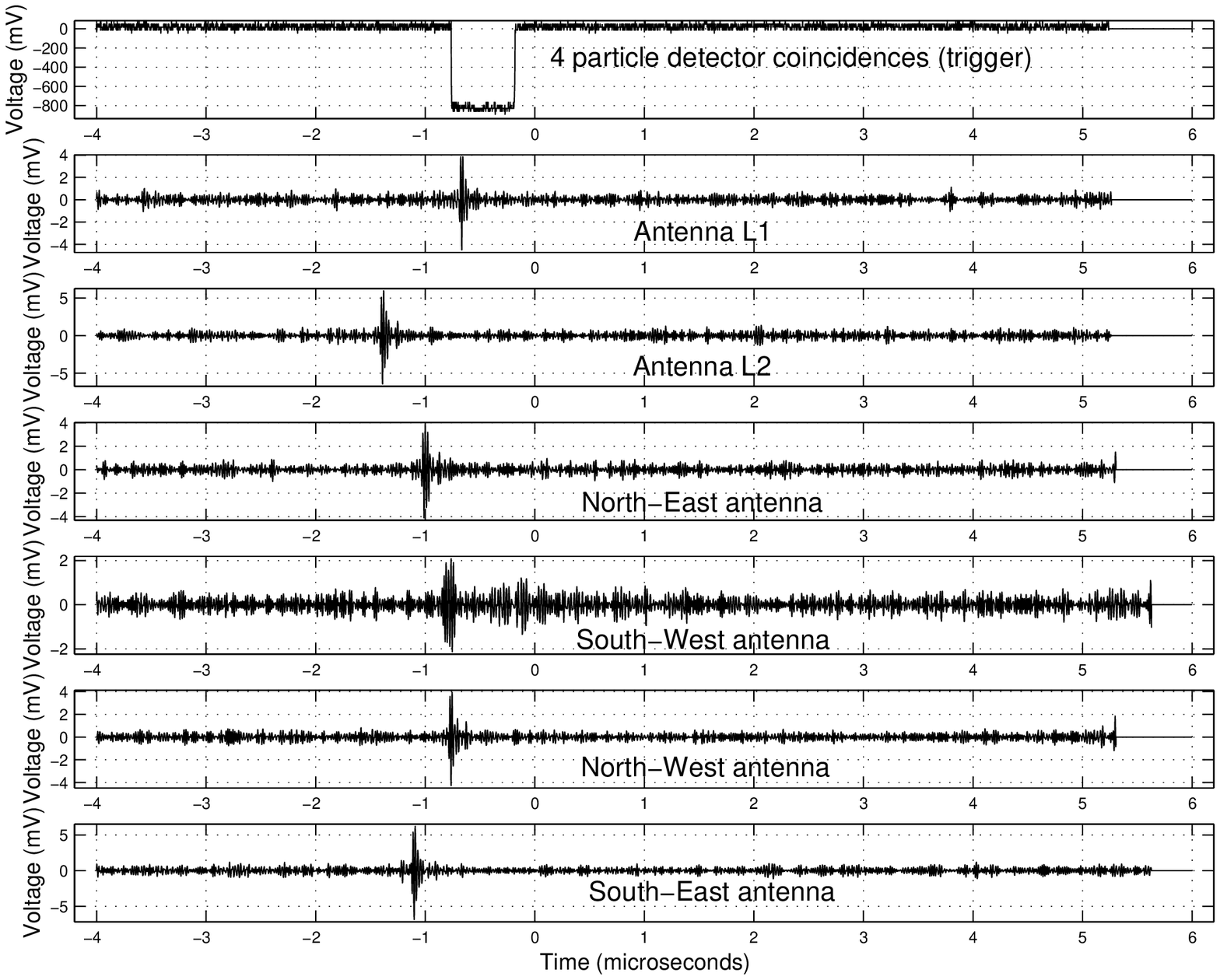,width=7.5cm}}
\vspace*{8pt}
\caption{Filtered antenna signals obtained for a cosmic ray event triggered by the particle detectors.\label{fig:r248e1782}}
\end{minipage}
\end{figure}

In the second phase operating since mid 2004, the above setup (see Fig. \ref{fig:setup2})
has been completed  with four double plastic scintillators \cite{Station} placed at the corner 
of the DAM array
 ($\simeq 100*100 m ^2$).
The trigger of the experiment is made of the four particle detectors in coincidence, resulting  
on an event rate of 0.8 event/mn. All the antennas 
have now the same role and are passband filtered (24-82 MHz) in order to increase the signal to noise ratio.

The observation of coincident events on antennas and charged particle detectors (Fig. \ref{fig:r248e1782}) 
demonstrates the association of antenna transient signals with
 the occurrence of extensive air showers. 
This unambiguous evidence of  radio 
signals through the simultaneous detection of shower particles will allow, 
for the first time, the characterisation of the shape and amplitude of air showers associated radio pulses.
A preliminary event rate of 1/(8 hours) is observed with antenna multiplicity ranging from 3 to 6.

From the corresponding deposited energy distribution in scintillators, one can infer the location
 of the air shower core. The time delays between the particle detectors 
allow the reconstruction of the shower axis. From these information, impact parameter effects
 can be studied especially those related to non vertical showers.
The latter are expected \cite{Gousset} to generate amplitude and shape field variations which
 will better show up in large atmosphere volumes accessible with radiodetection method. 
Purposely, 5 antennas will be installed (up to 400 m from the DAM) on a east-west line crossing
 the existing array.

Two effects, namely Cerenkov emission and the classical 
far field, contribute to the radio emission of a shower \cite{Gousset}. 
The line will also allow to study their respective influences
and asseses the interest for designing a larger antenna array dedicated to 
Ultra High Energy Cosmic Rays.


\end{document}